\documentstyle[12pt]{article}
\documentstyle{fleqn}
\begin{document}
\vfill\eject
\begin{flushright}
OU-HET 198
\linebreak
May  31, 1995
\end{flushright}
\vskip2cm
\begin{center}
{\large
{\bf  Effective Topological Theory for
Gravitational Anyon Scatterings at Ultra-High
Energies
}}
\end{center}
\vskip2cm
\begin{center}
{\large
Takahiro KUBOTA and
Hiroyuki TAKASHINO
}
\end{center}
\vskip1cm
\begin{center}
{\it Department  of Physics,
  Osaka University,
\linebreak
Toyonaka,
Osaka 560, Japan}
\end{center}
\vskip1cm
\begin{abstract}
The idea of the effective topological theory
for high-energy scattering proposed by H. and
E.  Verlinde is applied to the $(2+1)$ dimensional
gravity with Einstein action  plus Chern-Simons terms.
The calculational steps in the topological description
are compared with the eikonal approximation. It is
shown that the Lagrangian of the effective topological
theory turns out to vanish except for boundary
terms.
\vskip1cm

\end{abstract}
\pagebreak
\begin{center}
\large
{\bf \S 1. Introduction}
\end{center}

It has sometimes happened in  physics that  very
peculiar phenomena show up under  extreme conditions,
{\it e.g.},  asymptotic freedom in high energy QCD
scatterings, superconductivity at low temperature,
 strong gravitational forces near black holes and so
forth. There are therefore good reasons  to feel free
to do ``Gedanken Experiment" to uncover  deep
structures of theories at our hands.
As one of such Gedanken Experiments, there have been
several interesting  attempts $^{1)-4)}$
of inspecting  Planckian
energy scatterings. This  is expected   to serve as a
tool of inspecting quantum aspects of gravitational
theories, in particular, in the framework of
string theories.

There have been principally three methods to study the
Planckian energy scattering, namely, (i) the eikonal
approximation $^{1), 3)}$,
(ii) the shock wave method $^{2), 5)}$ and (iii)
 effective topological theory
proposed by H. and E.  Verlinde $^{6)}$.
The field-theoretical formulation  of the eikonal
approximation has been known $^{7), 8)}$ for long
time, while the shock wave method has
 come to our concern rather recently. It has been known
that high-energy small-angle scattering amplitudes
computed in the eikonal and shock wave methods
always agree. The  third method of Verlindes on the
other hand is motivated in such a way that the  most
dominant terms contributing to the leading eikonal
approximation are separated from the outset on the
Lagrangian level. A very peculiar fact in four
dimensional Einstein  gravity at the Planckian
energy is that the obtained effective Lagrangian
turns out to be topological $^{9)}$, {\it i.e.},
 being expressed as a BRST-exact
form up to surface integrals.  Kabat and Ortiz $^{10)}$
examined the effective topological method, comparing
it with the other methods in four dimensional Einstein
gravity.

Now that we have various methods of producing the same
leading approximation, an imminent issue is how to go
beyond the leading approximation
 by including non-leading terms. This is
important because several  interesting quantum effects
could emerge there.  There have been a lot of works
in this direction  in the first method $^{11), 12)}$.
As far as  Verlindes' topological method is concerned,
however, it does not look straightfoward
 to improve their approximation method $^{11)}$.

In the present paper,  we do not try  to go beyond the
leading approximation in Verlindes' method, but rather we
would like to reexamine the calculational  steps of
their method by  paying  particular attentions to the
connection between Verlindes' and the conventional
eikonal methods. We will apply the Verlindes' method to
the (2+1)-dimensional Einstein action
supplemented by Chern-Simons terms.
Although this gravitational theory is a theoretical
laboratory, it  is interesting in its own right
 since it produces fractional spin and statistics
$^{13)}$ as in the vector Chern-Simons theories
$^{14), 15)}$.
We will show that the (2+1)-dimensional
Einstein-Chern-Simons gravity is in fact described
effectively at high energies by a  topological field
theory, whose Lagrangian  turns out to vanish
except for boundary terms.

\vskip1cm
\begin{center}
\large
{\bf \S 2. The Gravitational Anyon}
\end{center}

As a starting point, let us begin with the
 total action
\begin{equation}
S=S_{E}+S_{CS}+S_{matter},
\end{equation}
where the Einstein and Chern-Simons actions are given
respectively by
\begin{equation}
S_{E}=\frac{1}{2 \kappa ^{2}}\int d^{3}x
\sqrt{g^{(3)}}
{  R^{(3)}},
\end{equation}
\begin{equation}
S_{CS}=\frac{1}{4\kappa ^{2}\mu}\int d^{3}x
\epsilon ^{\lambda
\mu \nu }\Gamma ^{\sigma }_{\lambda \rho}
(\partial _{\mu
 }\Gamma ^{\rho}_{\sigma \nu}+\frac{2}{3}\Gamma
^{\rho}_{\mu \tau}\Gamma ^{\tau}_{\nu \sigma}).
\end{equation}
The mass dimensions of the gravitational
constant $\kappa ^{2}$ in three space-time
dimensions is ${\rm dim}[\kappa ^{2}]=-1 $.

The (2+1)-dimensional gravity without the Chern-Simons
term has been a subject of intensive research in the
last decade $^{16)}$.  This theory is peculiar in that
there is no graviton and
that space-time is flat outside sources. Dynamics is
determined by global geometry. Scattering problems
in this theory was also investigated in detail $^{17)}$.
Inclusion of the Chern-Simons term renders  the theory
``topologically massive" with propagating modes $^{18)}$.
The Chern-Simons action contains
terms of third derivatives
with respect to the metric, and one may expect that
these terms would be of importance for high-energy
scatterings.  This is in contrast with the vector
 Chern-Simons
action where only first-derivative terms are contained.

For the sake of simplicity, we  consider the matter
action  consisting of only a scalar field interacting
with gravitational field.
\begin{eqnarray}
S_{matter}&=&\int d^{3}x\sqrt{g^{(3)}}(g^{(3)
\mu \nu}\partial _{\mu}\phi  ^{\dag}
 \partial _{\nu}\phi -
m^{2}\phi ^{\dag}\phi)         \nonumber  \\
&=& \int d^{3}x [\eta ^{\mu \nu}
\partial _{\mu }\phi ^{\dag}\partial _{\nu}\phi -
m^{2}\phi ^{\dag}\phi    \nonumber    \\
& & +h^{\mu \nu}\{-\partial _{\mu}\phi ^{\dag}\partial
 _\nu \phi +\frac{1}{2}\eta_{\mu \nu}(\partial ^{\lambda}
\phi ^{\dag}\partial _{\lambda} \phi-m^{2}\phi ^{\dag}
\phi)\}
  ]+{\cal O}(h^{2}).
\end{eqnarray}
Here we have expanded the metric around the
 flat space-time by setting
\begin{equation}
g_{\mu \nu}^{(3)}=\eta _{\mu \nu}+h_{\mu \nu},
\hskip1cm
\eta _{\mu \nu }=\rm{diag}(1,-1,-1).
\end{equation}
The higher-order terms with respect to fluctuations
$h_{\mu \nu}$ are not considered in the leading eikonal
approximation.

The Chern-Simons interaction (3) is known to produce
fractional statistics $^{13)}$,
which is in complete analogy
with the case of gauge interactions $^{14), 15)}$.
The emergence of the fractional statistics
is ascribed to the fact
that a non-trivial phase shows up
when  two point-particles
 are interchanged adiabatically. A simple way to
 evaluate  the phase is to look at the two-body
 static potential, or more generally the two-point
 function of $h_{\mu \nu}$
\begin{equation}
\Delta _{\mu \nu},_{\lambda \rho}(x,y)=\frac{-i}
{8\kappa ^{2}}<h_{\mu \nu } (x)h_{\lambda \rho }(y)>.
\end{equation}

Now as a preliminary calculation
for later use, we present
here the propagator when we take
the gauge fixing condition
specified by
\begin{equation}
{\cal L} _{g.f.}=-\frac{1}{4\kappa ^{2}\xi
 }(\partial _{\lambda}
h^{\lambda \mu}-\frac{1}{2}\partial ^{\mu}h_{\lambda}
^{\lambda})^{2}.
\end{equation}
(The propagator in the Landau  gauge has been
given in Ref. 18).) The
defining equation of the two-point
function becomes
\begin{eqnarray}
[A^{\mu \nu \lambda \rho}\partial ^{2}
& + & \frac{1}{2\mu}B^{\mu \nu  \lambda \rho}
-(1-\frac{1}{\xi})C^{\mu \nu \lambda \rho}
+2(1-\frac{1}{\xi}) D^{\mu \nu \lambda \rho}  \nonumber
 \\
& - & (1-\frac{1}{\xi})\eta ^{\mu \nu}\eta
^{\lambda \rho}
\partial ^{2}]
\Delta  _{\lambda \rho ,\sigma \tau }(x,y)
=\frac{1}{2}(\eta ^{\mu}_{\sigma}\eta ^{\nu}_{\tau }
+\eta ^{\mu}_{\tau }\eta ^{\nu}_{\sigma})
\delta ^{3}(x-y),
\end{eqnarray}
where we have introduced the following notations
\begin{eqnarray}
A^{\mu \nu \lambda \rho}& = & \eta ^{\mu \lambda}
\eta ^{\nu \rho}+\eta ^{\mu \rho}\eta ^{\nu \lambda}
-\eta ^{\mu \nu}\eta ^{\lambda \rho},  \\
B^{\mu \nu \lambda \rho}& = &
\epsilon ^{\nu \sigma \rho}
(\eta ^{\mu \lambda}\partial ^{2}-\partial ^{\mu}\partial
^{\lambda})\partial _{\sigma}
+ \epsilon ^{\mu \sigma \rho}
(\eta ^{\nu \lambda}\partial ^{2}-\partial ^{\nu}\partial
^{\lambda})\partial _{\sigma}
+(\lambda  \longleftrightarrow  \rho),    \\
C^{\mu \nu \lambda \rho} & = & \eta ^{\nu \rho}\partial
^{\mu}\partial ^{ \lambda}+\eta ^{\mu \rho}\partial
^{\nu }\partial ^{\lambda}+(\lambda \longleftrightarrow
\rho),                   \\
D^{\mu \nu \lambda \rho}& = & \eta ^{\lambda \rho}
\partial ^{\mu }\partial ^{\nu}+\eta ^{\mu \nu}
\partial ^{\lambda}\partial ^{\rho}.
\end{eqnarray}
The formal solution to (8) is given by
\begin{eqnarray}
\Delta _{\lambda \rho, \sigma \tau}(x,y)& = &
[
\frac{1}{4}A_{\lambda \rho \sigma \tau}
(\frac{1}{\partial ^{2}}-\frac{1}{\partial ^{2}+\mu ^{2}})
-\frac{\mu}{8}B_{\lambda \rho \sigma \tau}
\frac{1}{(\partial ^{2}+\mu ^{2})(\partial
^{2})^{2}}      \nonumber           \\
& &
+\frac{1}{4}C_{\lambda \rho \sigma \tau}
\{\frac{1}{(\partial ^{2}+\mu ^{2})\partial ^{2}}
+(\xi -1)\frac{1}{(\partial ^{2})^{2}}\} \nonumber  \\
& &
-\frac{1}{4}D_{\lambda \rho \sigma \tau}\frac{1}
{(\partial ^{2}+\mu ^{2})\partial ^{2}}
-\frac{1}{4}\frac{1}{(\partial ^{2}+\mu ^{2})
(\partial ^{2})^{2}}
\partial _{\lambda }\partial _{\rho}\partial _{\sigma}
\partial _{\tau}               \nonumber   \\
& &
-\frac{1}{4}\eta _{\lambda \rho}\eta _{\sigma \tau}
\frac{1}{\partial ^{2}}
]\delta ^{3}(x-y).
\end{eqnarray}

The anyonic interaction comes
from the second term in (13).
In fact for the static case, it produces an interaction
\begin{eqnarray}
& &-\frac{\mu}{8}B_{000i}\frac{1}{(-{\bf \nabla}^{2}
+\mu ^{2})(-{\bf \nabla}^{2})^{2}}\delta ^{2}({\bf x}-
{\bf y}) \delta (x^{0}-y^{0})
       \nonumber \\
& &=\frac{1}{8\pi}\epsilon_{0ij}\frac{x^{j}-y^{j}}
{\vert {\bf x}-{\bf y}\vert}\{K'_{0}(\mu \vert {\bf x}
-{\bf y}\vert)+\frac{1}{\mu \vert {\bf x}-
{\bf y}\vert}\}
\delta (x^{0}-y^{0}).
             \nonumber \\
\end{eqnarray}
Here $K_{0}$ is the modified Bessel function.
The two-body potential is obtained by multiplying
 (14) by the relative velocity $\frac{d}{dt}( x^{i}-
 y^{i})$   .
As
$\vert {\bf x}-{\bf y}\vert \rightarrow \infty $,
the modified Bessel function decreases exponentially,
and the second term in the
brackets in (14) is more dominant.
This term induces a phase proportional to
$\kappa ^{2}m^{2}/\mu$ as
the two particles are interchanged.

\vskip1cm
\begin{center}
\large
{\bf \S 3. The Eikonal Approximation}
\end{center}

In the following we summarize the eikonal approximation
 method in the gravitational theory (1). Some  of the
contents in this section was implicitly stated in Ref.
19), but will be given below for the sake of comparison
with the topological field theory.
 The technique of eikonal
approximation  has been well-known
and  explained  in quantum mechanics textbooks.
The analogous calculations in
quantum field theories have also been studied in the
sixties. Roughly speaking, the field theoretical
eikonal approximation amounts to summing up an infinite
number of Feynman diagrams of exchange type (Fig. 1)
without worring about self-energies or vertex
renormalzation effects, either.

Hereafter we will use the formulation
proposed by  Abarbanel and Itzykson $^{7)}$.
The linearized gravitational interaction in Eq. (4)
 motivates us to consider an operator

\begin{equation}
A(X,P)=P_{\mu}\{h^{\mu \nu}(X)-\frac{1}{2}\eta
^{\mu \nu}h^{\lambda}_{\lambda}(X)\}P_{\nu}
+\frac{1}{2}m^{2}h^{\lambda}_{\lambda}(X),
\end{equation}
where $X_{\mu}$ and $P_{\nu}$ are assumed to
satisfy formally the usual commutation relation
$
[X_{\mu},  P_{\nu}]=-i\eta _{\mu \nu}
$.

Each of the two particles in Fig. 1 is propagating
while emitting and/or absorbing
virtual gravitons due to the
linearized interactions.   The evaluation
of the propagator is  facilitated by making
use of the formula

\begin{eqnarray}
& &\lim _{P^{2}\rightarrow m^{2}}
(P^{2}-m^{2})\frac{1}{P^{2}-m^{2}-A(X,P)+i\epsilon}
(P^{2}-m^{2}) \nonumber   \\
& &=T\exp \{-i\int _{0}^{\infty}d\tau
A(X+2\tau P,P)\}A(X,P).
\end{eqnarray}
Here $T$ is the ordering according to the
variable $\tau $.
The matter propagator in Fig. 1 may be obtained just
by sandwitching the above formula by initial and final
states.

The Feynman diagrams in Fig. 1 are  evaluated by
connecting the gravitational lines by the
propagator (6). The combinatorial factors are such
that the invariant
amplitude is expressed by the following exponentiated
formula

\begin{eqnarray}
& &\frac{-i}{(2\pi)^{3}}
{\cal T}(s,t)\delta^{3}(p_{2}+p_{2}'
-p_{1}-p_{1}')   \nonumber  \\
&=& \exp\{8\kappa ^{2}i
\int \int d^{3}yd^{3}y'\frac{\delta}{\delta
h_{\mu \nu}(y)}\Delta _{\mu \nu ,\lambda \rho}(y,y')
\frac{\delta}{\delta h'_{\lambda \rho}(y')}
\}               \nonumber  \\
& &\times <p_{2}\vert T\exp\{-i\int
_{0}^{\infty}d\tau A(X+2\tau
P, P)\}A(X, P)\vert p_{1}>   \nonumber  \\
& &\times <p_{2}'\vert
T\exp\{-i\int _{0}^{\infty}d\tau 'A'(X+2
\tau 'P, P)\}A'(X, P) \vert p_{1}'>   \vert _{h=h'=0}  .
\end{eqnarray}
Here $s$ and $t$ are the usual Mandelstam variables.

To get the small angle scattering amplitude, all we
have to do is to replace the operator $P$
by the average
of the initial and final momenta {\it i.e.},
$p=\frac{1}{2}(p_{1}+p_{2})$,
$p'=\frac{1}{2}(p_{1}'+p_{2}')$.
After this manipulation we arrive at the eikonal
amplitude (to be denoted hereafter by
${\cal T}_{E}$),

\begin{eqnarray}
{\cal T}_{E}(s, t)&=&-2i{\bar s}
\int db \exp \{ ib\cdot (p_{2}-p_{1})\}
         \nonumber \\
& &\times [\exp \{-8\kappa ^{2}i
\int _{-\infty}^{\infty}d\tau
\int _{-\infty}^{\infty}d\tau '
(p^{\mu }p^{\nu}
-\frac{1}{2}\eta ^{\mu \nu}(p^{2}-m^{2}))
                       \nonumber \\
& & \times  \Delta _{\mu \nu ,
\lambda \rho}(b+2p\tau -2p'\tau ',
 0)
 (p'^{\lambda }p'^{\rho }-\frac{1}{2}\eta ^{\lambda
\rho}(p'^{2}-m^{2}))
\}
            \nonumber \\
& &-1].
\end{eqnarray}
Here $b$ is a vector satisfying the orthogonality
$b\cdot p=b\cdot p'=0$
and we have also introduced a notation

\begin{equation}
{\bar s}=s \sqrt{1-\frac{4m^{2}-t}{s}}.
\end{equation}

Keeping terms up to linear in $t/s$ and $m^{2}/s$,
we get
\begin{equation}
{\cal T}_{E}(s,t)\sim
-2i{\bar s}\int _{-\infty} ^{+\infty} db
\exp (\pm i\sqrt{-t}b)[\exp
\{\frac{i}{2}\kappa ^{2}{\bar s}
\Delta (b)\}-1],
\end{equation}
where
\begin{equation}
\Delta (b)=\frac{1}{2}\vert b \vert +\frac{1}{2\mu}
\epsilon (b)+\frac{1}{2\mu}\exp \{-\mu \vert b \vert \}
-\frac{1}{2\mu}\epsilon (b)\exp \{-\mu \vert b \vert \}.
\end{equation}
Here $\pm $ in the exponent in (20)
indicates that, if the
deflection angle  after scattering
is positive (negative)
in the center of mass system,
then we should take minus (plus)
sign. Note that the gauge parameter
dependence disappeared
in this approximation. We can see
easily that the inequality
$\Delta (b)\not= \Delta(-b)$
 produces aymmetry of the amplitudes
with respect to the
deflection angle.
The  asymmetric phase in (20) is analogous to  the
Aharonov-Bohm effect discussed
in sec. 2. There is, however, slight
difference in that, while the phase of
the fractional statistics
is proportional to $\kappa ^{2}m^{2}/\mu $ ,
 the counterpart
in (20) is proportional to $\kappa ^{2}{\bar s}/\mu $.
Incidentally, note that $\Delta (b)$ satisfies the
differential equation
\begin{equation}
(\frac{d ^{2}}{d b ^{2}}
-\frac{1}{\mu}\frac{d ^{3}}{d b ^{3}}
)\Delta (b)=\delta (b).
\end{equation}

\vskip1cm
\begin{center}
\large
{\bf \S 4. A Qualitative Analysis  towards
Topological Description}
\end{center}

Here we digress a little while to discuss the eikonal
formula (20), and its implications. In passing from (18)
to (20), we have contracted Lorentz indices term by
term. The calculation is straightforward but rather
tedious. There is, however,
more direct way to reach (20).

Let us  make use of the light-cone variables, i.e.,
$x^{\pm}=(x^{0}\pm x^{1})/\sqrt {2},$
$x^{\perp}=x^{2}.$
We will call the $\pm$-direction longitudinal, while
$\perp $-direction transverse.
Suppose that the momenta $p^{\mu}$ and $p'^{\mu}$
lie in the longitudinal direction at the ultra-high
energy. The most dominant component of
$p^{\mu}$ ($p'^{\mu}$) is $p^{+}$ ($p'^{-}$).
The summation over the Lorentz indices is then very
much simplified, and the only contribution is reduced to
the longitudinal component
$\Delta _{++,--}(b+2p\tau -2p'\tau ', 0)$.

The dynamics under the eikonal approximation is that the
particle interactions occur at short distance in the
longitudinal direction,  i.e., $2p\tau -2p'\tau '\sim 0$,
which is in contrast with
the rather large distance interaction
in the transverse component.
Considering these facts we may
put the formal solution of
the Green's function (13) in the
following manner.
The derivatives in the transverse
direction may be important
and must be kept throughout. Those in the longitudinal
direction, on the other hand, pick up small corrections
in the ultra-high energy scatterings
and may be discarded in our
problem. These approximation amounts to the following
\begin{eqnarray}
\Delta _{++,--}(x,y)  &\sim &[\frac{1}
{4}A_{++--}(\frac{1}{-\partial
_{\perp}^{2}}-\frac{1}{-\partial _{\perp} ^{2}+\mu ^{2}})
                          \nonumber \\
& &-\frac{\mu }{8}B_{++--}\frac{1}{(-\partial _{\perp}
 ^{2}+\mu ^{2})(-\partial _{\perp}^{2}
 )^{2}}]\delta ^{3}(x-y)
                   \nonumber \\
&=&-\frac{1}{2}(\partial _{\perp} ^{2}-\frac{1}{\mu}
\partial _{\perp}^{3})^{-1}\delta ^{3}(x-y).
\end{eqnarray}
Here we have put
$A_{++--}=2$
and
$B_{++--}\sim 4\partial _{\perp}^{3}$.
We immediately notice that the same differential operator
 has emerged in (23) as in (22).
In other words, the Green function
is replaced by
\begin{equation}
\Delta _{++,--}(b+2p\tau -2p'\tau ', 0)\sim
-\frac{1}{2}\Delta (b)\delta ^{2}(2p\tau -2p'\tau ')
\end{equation}
and thereby we can easily jump  from (18) to (20).

The separation of the dynamics into longitudinal and
transverse sectors suggests a new way of looking at the
eikonal approximation. It is all
in the ``transverse" Green
function $\Delta (b)$ where the dynamics of the
scatterings is contained.  Apparently it is possible
to extract $\Delta (b)$ on the Lagrangian level
by sorting out the most dominant ones
 among the kinetic terms. It is, however, not quite
obvious whether the effective Lagrangian after
sorting out is a topological field theory.  This is
what we would like to see next and will show that the
effective Lagrangian is non-vanishing only on
boundaries.

\vskip1cm
\begin{center}
\large
{\bf \S 5. Effective Topological Field Theories}
\end{center}

Now we are in a position to discuss the same scattering
problem in the effective topological method proposed by
Verlindes $^{6)}$.  We would like to see to what extent
the Verlindes' idea  works for the action (1).
As we have seen, the
factorization of the dynamics along the longitudinal and
transverse
direction simplifies the problem considerably, and
we are led to take the following gauge choice
\begin{equation}
g_{\mu \nu}^{(3)}=
\left (
\begin{array}{cc}
g_{\alpha \beta} &  \begin{array}{c}
                    0 \\
                    0
                    \end{array}
\\
     \begin{array}{cc}
      0 & 0 \\
     \end{array} &  h
\end{array}
\right ),
\hskip1cm
(\alpha, \beta =0, 1).
\end{equation}
It is assumed that $h$ is space-time independent  and
is just a constant. Physical
quantities should not depend on
$h$. The ghost action in the case of the above
gauge choice becomes
\begin{equation}
S_{gh}=
\int d^{3}x\sqrt{g^{(3)}}\{b_{2\alpha}(\nabla ^{2}
c^{\alpha}+\nabla ^{\alpha}c^{2})+2b_{22}
\nabla ^{2}c^{2}\}.
\end{equation}

Previously, in order to reach Eq. (20) we have neglected
the derivatives along the longitudinal direction and
have kept only those in the transverse one in the
propagators.
This procedure may be achieved equivalently on the
Lagrangian level by considering
scaling behaviors under the change of the metric
\begin{equation}
g_{\alpha \beta}\rightarrow l_{\parallel}^{2}g_{\alpha
\beta},
\hskip1cm
h\rightarrow l_{\perp}^{2}h.
\end{equation}
Here $l_{\parallel}$ and $l_{\perp}$ are the typical
length scales characterizing the longitudinal
 and transversal dynamics, respectively.
In order to see the meaning of the scaling
properties under (27) , let us separate the
Einstein, Chern-Simons and ghost actions
into two parts, {\it i.e.,} longitudinal and
transverse ones,
\begin{equation}
S_{E}=S_{E \parallel}+S_{E \perp},  \hskip0.5cm
S_{CS }=S_{CS \parallel}+S_{CS \perp},\hskip0.5cm
S_{gh}=S_{gh \parallel}+S_{gh \perp}.
\end{equation}
Each term in (28) is defined by
\begin{eqnarray}
S_{E \parallel}&=&\frac{1}{8\kappa ^{2}}
\int d^{3}x\sqrt{-g}\sqrt{-h}
h^{-1}(\partial _{2}g_{\alpha \beta})
(\partial _{2}g_{\gamma \delta})(g^{\alpha \beta }
g^{\gamma \delta}-g^{\alpha \gamma}g^{\beta \delta}
),            \\
S_{E \perp}&=&\frac{1}{2\kappa ^{2}}\int d^{3}x\sqrt{-g}
\sqrt{-h}R_{g},              \\
S_{CS \parallel}&=& \frac{1}{4\kappa ^{2}\mu}
\int d^{3}x \epsilon ^{\alpha \beta 2}\{
 \Gamma _{\beta 2}^{\gamma}(\partial _{2}\Gamma
_{\gamma \alpha}^{2})-\Gamma _{\alpha \gamma}^{2}
(\partial _{2}\Gamma _{\beta 2}^{\gamma})\},
                              \\
S_{CS \perp}&=&\frac{1}{4\kappa ^{2}
\mu}\int d^{3}x \epsilon
^{\alpha \beta 2}\{\Gamma _{\delta \beta}^{\gamma}
(\partial _{2}\Gamma _{\gamma \alpha }^{\delta})
+\partial _{\beta}(\Gamma _{\alpha \delta}^{\gamma}
\Gamma _{\gamma 2}^{\delta})   \},    \\
S_{gh \parallel}&=&\int d^{3}x \sqrt {-g}
\sqrt{-h}h^{-1}\{b_{2\alpha}(\partial _{2}c^{\alpha})
+2b_{22}(\partial _{2}c^{2})\},    \\
S_{gh \perp}&=&-\int d^{3}x\sqrt{-g}
\sqrt{-h}g^{\alpha \beta}
(\nabla _{\alpha}b_{2\beta} )c^{2}.          \\
\nonumber
\end{eqnarray}
Here $R_{g}$ in (30) denotes the two-dimensional scalar
 curvature  associated with $g_{\alpha \beta}$.

As the notations show, the longitudinal and transverse
parts of the actions  are tansformed distinctively
under (27) {\it i.e.,}
\begin{equation}
S_{E \parallel}\rightarrow \frac{l_{\parallel}^{2}}
{l_{\perp}}S_{E \parallel},
\hskip1cm
S_{E \perp}\rightarrow l_{\perp}S_{E \perp},
\end{equation}
\begin{equation}
S_{CS \parallel}\rightarrow (\frac{l_{\parallel}}
{l_{\perp}})^{2}S_{CS \parallel},
\hskip1cm
S_{CS \perp}\rightarrow S_{CS \perp},
\end{equation}
\begin{equation}
S_{gh \parallel}\rightarrow \frac{l_{\parallel} ^{2}}
{l_{\perp}}S_{gh \parallel},
\hskip1cm
S_{gh \perp}\rightarrow l_{\perp}S_{gh \perp}.
\end{equation}
The scaling behavior (36) differs from those in (35)
 and (37).   We should, however, recall the
fact that $S_{CS}$ contains
the  mass parameter $\mu $ in front.
If $\mu $ is of the same order as $1/l_{\perp}$, then
the behavior (36) may be regarded as the same as (35)
 and (37).

In the high energy limit, the scattering dynamics
is confined in the short distance
region in the longitudinal
 direction.  This means that the
path integral region over the metric
corresponding to
 $l_{\parallel} \ll l_{\perp}$
is most important, and  $S_{E \perp}$, $S_{CS \perp}$,
and $S_{gh \perp}$ may be treated
classically. We will look for the local minimum
of $S_{E \perp}$, $S_{CS \perp}$ and $S_{gh \perp}$
by varying the fields.
The stability condition
$\delta S_{E \perp}=0$ provides us
immediately with
\begin{equation}
g_{\alpha \beta}=\partial _{\alpha}X^{a}
\partial _{\beta}X_{a}.
\end{equation}
In other words, the conformal mode is not important
and the metric is parametrized only by the two modes
$X^{a}$,  $(a=1,2)$.
Another stability condition of the ghost part $\delta
S_{gh \perp}=0$ turns out to be
$\nabla _{\alpha }b_{2\beta}=0 $
and we may set
\begin{equation}
b_{2 \alpha}=\epsilon _{\alpha \beta 2}\partial ^{\beta}b.
\end{equation}
The local minimum of  $S_{CS \perp}$  may be found easily
by solving the equation $\partial _{2}\Gamma ^{\beta}
_{\alpha \gamma}=0$, or more simply by noting the relation
\begin{equation}
\delta S_{CS \perp}=\frac{1}{4\kappa ^{2}\mu}\int d^{3}x
\epsilon ^{\alpha \beta 2}\delta g_{\beta \gamma}\nabla
_{\alpha}\nabla ^{\delta}(\nabla _{\delta}V^{\gamma}-
\nabla ^{\gamma}V_{\delta})=0.
\end{equation}
Here we have introduced $V^{\alpha}$
defined by the relation
\begin{equation}
\partial _{2}X^{a}+V^{\alpha}\partial _{\alpha}X^{a}=0.
\end{equation}
The most dominant configuration of $X^{a}$ is realized
by imposing
\begin{equation}
\nabla _{\alpha}V_{\beta}-\nabla _{\beta}V_{\alpha}=0.
\end{equation}
By putting all these conditions into the longitudinal part
of the action, we arrive at

\begin{equation}
S_{E \parallel}=\frac{1}{2\kappa ^{2}}\int d^{3}x\sqrt{-h}
h^{-1}\epsilon  _{ab}\epsilon ^{\alpha \beta 2}
\partial _{\alpha }(\partial _{2}X^{a}\partial _{\beta}
\partial _{2}X^{b}),
\end{equation}

\begin{equation}
S_{CS \parallel}=\frac{1}{2\kappa ^{2}\mu}\int
d^{3}x  h ^{-1} \epsilon ^{\alpha \beta 2}
\partial _{\alpha}(\partial _{2}X^{a}\partial _{\beta}
\partial _{2}\partial _{2}X_{a}).
\end{equation}
Note that these are both in the
form of total divergence.

Finally let us come to the interaction of the scalar
 field and gravity.
The interaction part is also expressed
as a surface integral
\begin{equation}
S_{int}=\int d^{3}x \sqrt{-h}\partial ^{\alpha}(P_{a,\alpha}
X^{a}).
\end{equation}
Here $P_{a,\alpha}=T_{\alpha \beta}\partial ^{\beta}X_{a}$
is the momentum flow defined
in terms of the energy momentum
 tensor $T_{\alpha \beta}$.
To sum up, the effective Lagrangians in (43), (44) and (45)
are all expressed in the form of
total divergences and hence
topological.

It has been observed by Verlindes in four dimensional
gravity that the Jacobian associated with the change of
the path integral variables $g_{\alpha \beta}\rightarrow
X^{a}$ is exactly cancelled by another Jacobian due to
the change of the antighost. This cancellation, however,
does not occur in the three dimensional case, because the
Jacobian due to the change $b_{a\alpha}\rightarrow b$
 is only one half of the bosonic counterpart.  This
incomplete cancellation of the Jacobians would have
to be given  due consideration if we would go beyond the
leading approximation.  As far as we restrict ourselves
to the first approximation, however,
 the effect due to the Jacobian does not matter.

Another  remark is on  the work by Zeni $^{20)}$, who
applied the Verlindes' method to the Einstein gravity
in three dimensions.  Since Zeni did not have $S_{CS}$
 in his Lagrangian, the condition (42) did
not come from (40). Instead
he  considered the Gauss law constraints and has
obtained (42). Without imposing the constraints on the
Lagrangian level,
the effective theory would have been much more complicated.
In our case, on the other hand, the presence of $S_{CS}$
is crucial to get (42) and $S_{E \parallel}$  and $S_{CS
\parallel}$ have therefore become the form of
total divergence.

\vskip2cm
\begin{center}
\large
{\bf \S 6. Discussions}
\end{center}

In order to see what the scattering amplitudes look like
in the effective topological theory, we rewrite the
 effective  Lagrangian in the form of contour integrals

\begin{eqnarray}
& &S_{E \parallel}+S_{CS \parallel}+S_{int}
          \nonumber \\
& &=\frac{-1}{2\kappa ^{2}}
\oint d\sigma \int \sqrt {-h}h^{-1}
\frac{\partial X^{a}}{\partial \sigma}\{\epsilon _{ab}
(\partial _{2})^{2}+\frac{1}{\mu \sqrt{-h}}
\eta _{ab}(\partial _{2})^{3}\}X^{b}
                          \nonumber \\
& &+\oint d\sigma \int \sqrt{-h}\epsilon
 ^{\sigma \alpha 2}(P_{a,\alpha }X^{a})
                   \nonumber \\
\end{eqnarray}
The variable $\sigma $ parametrizes the boundary $C$
of the two-dimensional manifold $(x^{0}, x^{1})$
as shown in Fig. 2. The variable $X^{a}$ is now
regarded as a function of $\sigma$ and $x^{2}$
{\it i.e.,} $X^{a}=X^{a}(\sigma, x^{2})$.
The differential operator in the kinetic term of
$X^{a}$ is similar in form as those in Eqs. (22) and (23).
Thus the combined use of the scaling
argument and the semiclassical treatment of the
 transverse part of the action enables us to extract the
Green function (21) on the Lagrangian level.
It is in fact easy to see that the Green function (21)
shows up in the two-point function
\begin{equation}
<X^{+}(\sigma ,b)X^{-}(\sigma ' , b')>=\frac{-i
\kappa ^{2}}{2}
\Delta (-\sqrt{-h}(b-b'))\epsilon (\sigma -\sigma ').
\end{equation}

The path integral over $X^{a}$ of the action (46)
result in
\begin{eqnarray}
& &\exp \{\frac{i\kappa ^{2}}{2}\oint d\sigma
\oint d\sigma ' \int db \sqrt{-h}\int db' \sqrt{-h}
\epsilon ^{\sigma \alpha 2}\epsilon
^{\sigma ' \beta 2}P_{+,\alpha}(\sigma, b)
P_{-,\beta}(\sigma ', b')  \nonumber \\
& & \times \Delta (-\sqrt{-h}(b-b'))\epsilon
 (\sigma -\sigma ')\},
                           \nonumber \\
\end{eqnarray}
where we have used the fact that the Green functions
$<X^{+}X^{+}>$ and $<X^{-}X^{-}>$  both vanish.
At ultra-high energies, only the components
$P_{+,+}$ and $P_{-,-}$ survive the summation
in the above exponent
and the integrations of $P_{\pm, \pm}(\sigma, b)$ over
$\sigma $ and $b$ give us the incoming and outgoing
momenta of each particle. Suppose that the incident
 and outgoing particles have some particular relative
impact parameters, then the exponentiated form in (48)
is exactly the same as the integrand in (20). It is
thus possible in the effective topological method to
set up a calculational scheme for scattering
amplitudes.  Namely scattering amplitudes are given
by correlation functions of ``vertex operator"
$\exp \{iS_{eff}\}$ and Verlindes' program is now
fulfilled for the (2+1)-dimensional
Einstein-Chern-Simons gravity.

Finally a few remarks are in order with regard to previous
 related works in literatures. Deser, McCarthy and Steif
$^{19)}$ studied the same scattering problem in the
shock-wave and eikonal methods. They started with the
metric of the Aichelburg-Sexl type
\begin{equation}
ds^{2}=2dx^{+}dx^{-}-(dx^{\perp})^{2}-2F(x^{-},
x^{\perp})(dx^{-})^{2}.
\end{equation}
The equation satisfied by $F(x^{-}, x^{\perp})$ is of the
same form as that of $\Delta _{++,--}$ and they examined
ambiguity problems associated with the solution.  The same
problems also remain in the effective topological method;
If  one would start from the effective Lagrangian (46)
and tried to get the Green function
$<X^{+}(\sigma ,b)X^{-}(\sigma ', b')>$,
the conventional $i\epsilon $-prescription is no more
available and one would encounter the problem as to how
to fix the boundary conditions.  We are forced to go back
to the original Green function (13) to fix the boundary
conditions.  This fact is  unsatisfactory in the effective
topological method.

Comparison between the topological and the conventional
eikonal methods offers several implications as to the
sub-leading terms. Amati, Ciafaloni and Veneziano
$^{11)}$  studied sub-leading corrections to the eikonal
approximation in the four dimensional gravity and have
realized that the Verlindes' gauge choice has intrinsic
difficulties. The problem comes from particular metric
fluctuations in the off-diagonal part of the metric which
are not to be gauged away.
The same difficulties still remain
in the (2+1)-dimensional case and we do not have much to
 say about them. Furthermore since some of the sub-leading
terms are contained in the transverse part of the action,
there seems to be little chance that
the effective theory could be
a topological field theory at the sub-leading order.

\vfill\eject
\begin{center}
\large
{\bf Acknowledgements}
\end{center}
We would like to thank our colleagues at Osaka
University for their kind interest in the present
work.  This work was supported in part by the Grant
in Aid for Scientific Research from the Ministry of
Education, Science, and Culture (Grant No. 06640396).

\pagebreak
\begin{center}
\large
{\bf References}
\end{center}
\begin{description}
\item{1)}
D. Amati, M. Ciafaloni and G. Veneziano,
Phys. Lett. {\bf 197B} (1987), 81; Int.
J. Mod. Phys. {\bf A3}  (1988), 1615.
\item{2)}
't Hooft,
Phys. Lett. {\bf 198B} (1987), 61;
Nucl Phys. {\bf B304} (1988), 867;
  C. Klimcik,    Phys. Lett. {\bf B 208} (1988), 373.
\item{3)}
I. Muzinich and M. Soldate, Phys. Rev.  {\bf D37}
 (1988), 353.
\item{4)}
D.J. Gross and P.F. Mende, Phys. Lett. ; {\bf 197B}
  (1987), 129;
Nucl. Phys. {\bf B303} (1988), 407.
\item{5)}
P.C. Aichelburg and R.U. Sexl, Gen. Rel. Grav.
 {\bf 2} (1971), 303; T. Dray and G. 'tHooft,
Nucl. Phys. {\bf B253} (1985), 173.
\item{6)}
 H. Verlinde and E. Verlinde, Nucl. Phys.
{\bf B371}  (1992), 246.
\item{7)}
 H.D.I. Abarbanel and C. Itzykson, Phys. Rev. Lett.
 {\bf 23} (1969), 53; M. Levy and J. Sucher, Phys. Rev.
{\bf 186} (1969), 1656.
\item{8)} H. Cheng and T.T. Wu,  Phys. Rev. {\bf 186}
(1969), 1611.
\item{9)}
E. Witten, Commun. Math. Phys. {\bf 117} (1988),
353; {\bf 118} (1988), 411; {\bf 121} (1989),
351; Phys. Lett. {\bf 206B}  (1988), 601.
For a review see,
D. Birmingham, M. Blau, M. Rakowski and G. Thompson,
Phys. Reports {\bf 209C} (1991), 129.
\item{10)}
D. Kabat and M. Ortiz, Nucl. Phys. {\bf B 388}
(1992), 570.
See also R. Jackiw, D. Kabat and M. Ortiz, Phys.
Lett. {\bf B 277} (1992), 148.
\item{11)}
D. Amati, M. Ciafaloni and G. Veneziano, Nucl.
Phys. {\bf B 347}  (1990), 550;
{\bf  B403} (1993), 707; Phys. Lett. {\bf B289}
  (1992),  87;
\item{12)}
M. Fabbrichesi, R. Pettorino, G. Veneziano
and G.A. Vilkovisky, Nucl. Phys. {\bf B419} (1994),
 147.
\item{13)}
S. Deser, Phys. Rev. Lett. {\bf 64} (1990), 611;
S. Deser and J.G. McCarthy, Nucl.   Phys.  {\bf B344}
(1990), 747;
M. Reuter, Phys. Rev. {\bf D44} (1991), 1132.
\item{14)}
 J.M. Leinaas and J. Myrlheim, Nuovo Cimento
{\bf B37} (1977), 1;  F. Wilczek,   Phys. Rev.  Lett.
{\bf 48}  (1982),  1144;  {\bf 49} (1982), 957;
F. Wilczek and A. Zee, Phys. Rev. Lett. {\bf 51} (1983),
2250.
\item{15)}
  S. Forte,  Rev. Mod. Phys. {\bf 64}
 (1992), 193;  F. Wilceck,  {\it  Fractional
Statistics and Anyon Superconductivity}
(World Scientific Pub., 1990)
\item{16)}
 S. Deser, R. Jackiw and G. 'tHooft,
Ann. Phys. {\bf 152} (1984), 220.
\item{17)}
G. 'tHooft,  Commun. Math. Phys. {\bf 117} (1988), 685;
S. Deser and  R. Jackiw,   Commun. Math. Phys. {\bf 118}
 (1988),  495;
D. Lancaster and N. Sasakura,  Class. Quantum Grav.
 {\bf 8} (1991), 1481.
\item{18)}
S. Deser, R. Jackiw and S. Templeton,
 Ann. Phys. {\bf 140} (1982), 372; Phys. Rev. Lett.
{\bf 48} (1982), 975;
\item{19)}
S. Deser and A.R. Steif,  Class. Quantum Grav.
{\bf 9} (1992), 153;
S. Deser, J. McCarthy and A.R. Steif,
Nucl. Phys. {\bf B412} (1994), 305.
\item{20)}
M. Zeni,  Class. Quantum Grav.  {\bf 10} (1993),
905.

\pagebreak
\begin{center}
{\bf Figure Captions}
\end{center}
\item{Fig. 1}
Feynman diagrams summed up in the formula (17). The solid
lines denote the scalar particles, the dashed lines the
graviton exchange.
\item{Fig. 2} The contour $C$ of the
$\sigma $-integration in Eq.  (46) for
 the ultra-high energy scatterings
\end{description}
\end{document}